\documentclass[11pt,a4paper]{article}
\usepackage{graphicx}
\usepackage{amsmath}
\usepackage{cite}
\usepackage{verbatim} 
\usepackage[usenames]{xcolor}
\usepackage{url}
\usepackage{caption}

\newcommand{\bEq}{\begin{equation}}
\newcommand{\eEq}{\end{equation}}
\newcommand{\bEQ}[1]{\begin{equation} \begin{array}{#1}}
\newcommand{\eEQ}{\end{array} \end{equation}}

\newcounter{CorrectThis}

\begin{document}

\pagestyle{empty}

\null

\vfill

\begin{center}

{\Large  
{\bf Interface-plane parameterization
	
\vskip 0.2cm
for macroscopic grain boundary identification}
}\\

\vskip 1.0cm
A. Morawiec
\vskip 0.2cm 
{Institute of Metallurgy and Materials Science, 
Polish Academy of Sciences, \\ Krak{\'o}w, Poland.
}
\\
E-mail: nmmorawi@cyf-kr.edu.pl \\
Tel.: ++48--122952854, \ \ \  Fax: ++48--122952804 \\

 \end{center}

\vfill

\noindent
{\bf Abstract}
\\
The geometric state of a flat boundary is frequently described 
using the so-called macroscopic parameters. 
They are a principal tool for 
dealing with interfaces at the continuous scale.
The paper describes a new method for macroscopic identification of boundaries.
The proposed approach is based on Euler angles
representing orientations of the crystals. 
Two pairs of the angles are directly related 
to two vectors normal to the boundary plane 
in the crystal reference frames, and the new boundary representation 
can be viewed as a triplet composed of these vectors 
and the angle of rotation about the axis perpendicular to the plane.
The representation resembles the `interface-plane scheme', 
but unlike the latter, it is a proper parameterization. 
Basic practical aspects of the parameterization 
(such as equivalences due to symmetries, fundamental regions, 
uniform distribution of boundaries) are considered.
The parameterization is applied to examination of  
Bulatov-Reed-Kumar model of grain boundary energy
and reveals its previously unknown features. 
The proposed boundary identification method, 
apart from its use in numerical calculations, 
appeals to physical intuition.

\vskip 0.5cm


\vskip 0.0cm

\noindent
\hfill  \today

\newpage

\pagestyle{plain}

\noindent
\section{Introduction}

Quantities that identify boundary types and enable them to be distinguished
are essential for the study of grain boundaries.  
The basic method of describing a planar boundary is to give the misorientation 
between the crystals and the orientation of the boundary plane 
\cite{Goux_1961,Lange_1967,Fortes_1972,Goux_1974}.
From the atomistic viewpoint, the 
purely geometric macroscopic concept of a flat zero-thickness boundary is meaningless, 
but in practice, 
details of boundary structures and atomic-scale translations
are often inaccessible. 
In particular, 
only the macroscopic description is possible when 
experimental data on boundaries are limited 
to approximate boundary locations and grain orientations.
The macroscopic approximation, 
while avoiding the complexity that would arise 
if atomic positions or microscopic data were considered,
takes into account crystal anisotropy, 
which is important for explaining boundary properties.

The five real variables specifying grain misorientation 
and boundary plane are referred to as macroscopic 
boundary parameters. 
A parameterization must be an onto 
mapping, i.e., 
first, 
each set of parameters is assigned to exactly one boundary type,
and second, 
for every boundary configuration there exists a set of parameters assigned to it. 
The macroscopic boundary parameters 
work behind the scenes in software 
dealing with boundary data sets.
They are used to
characterize experimental results (e.g., \cite{Zhu_2022,Korolev_2022,Patel_2025}),  
in boundary simulations (e.g., \cite{zhang2022molecular,homer2022examination}),
in modeling of microstructures (e.g., \cite{Wei_2020,Murgas_2022}), and 
in many other 
studies on properties of boundary networks 
(e.g., \cite{Bizana_2023,Panzarino_2016}).

Numerous papers on grain boundaries emphasize the importance 
of crystal planes on both sides of the boundary.
Some approaches -- from Pumphray's `plane matching theory' \cite{Pumphrey_1972},
through the idea of `special boundary planes' \cite{Randle_1997,Randle_2006},
to the recently promoted 
`grain boundary inter-connection' \cite{Wang_2017,Wang_2018,Wang_2019} -- 
are focused on the crystal planes.
In such cases, the most appropriate boundary parameterization 
would be one that is based directly on parameters of the boundary planes.
Boundaries in bicrystals are often characterized by 
Miller indices of  abutting 
crystal 
planes. 
This description is incomplete,
but it brings to mind  
an attractive boundary parameterization.
The idea is to use 
parameters of the boundary plane with respect to reference frames of the two crystals 
and a fifth parameter -- the angle, say $\chi$, of a rotation about axis perpendicular to the plane.
The concept is intuitively simple, clear and compelling, 
but the practical problem of precisely defining such parameters has eluded researchers.
Wolf and Lutsko introduced the 'interface-plane scheme'
based on a triplet consisting of two vectors
representing the boundary plane in reference frames of the 
crystals on both sides of the boundary and a `twist' angle  \cite{Wolf_1989}.\footnote{See
also \cite{Takashima_2000} where relationships defining  the interface-plane scheme were completed and \cite{Morawiec_2013} for a more general 
approach to the boundary decomposition
into `twist' and `tilt'.}
However, the interface-plane scheme is not a proper 
parameterization 
because some physically distinct 
boundaries correspond to the same 
triplets \cite{Morawiec_2012}.

This paper describes five quantities, two pairs characterizing the boundary 
plane with respect to the crystal reference 
frames and an additional angle, 
which together constitute a proper parameterization. 
The parameters can be interpreted as Euler angles
related to orientations of the crystals 
or, 
more conveniently, they  
can be viewed as a triplet composed of two vectors identifying 
the orientation of the boundary plane in the crystals 
and the angle of rotation about an axis perpendicular to the plane.
This new boundary representation 
will be called the `interface-plane parameterization'.

The remainder of this paper is organized as follows. 
After recalling some basics and establishing the notation,
the five parameters are defined, and 
they are linked to unit vectors normal to the boundary plane 
in the reference frames of the crystals. 
Next, aspects important for practical applications 
of the interface-plane parameterization are presented:
sets of parameters equivalent due to symmetries 
and fundamental regions 
with representatives of the equivalence classes 
are considered, 
forms of interface-plane parameters 
of characteristic boundaries are described,
and a model of uniform boundary distribution
is briefly discussed.  
Before concluding, 
the new parameterization is applied to examination of 
a model of grain boundary energy. 
For simplicity, the considerations below are limited  
to the case of homophase interfaces 
between centrosymmetric crystals.

\section{Preliminaries}

\subsection{Ways to represent grain boundaries}

Practical methods of representing boundaries 
by the misorientation/plane pair take 
various
specific forms 
because there are many ways to describe misorientations
and plane orientations.
When the boundary plane is given in the reference frame of one of the crystals,
the description based on the misorientation and plane is `asymmetric'. 
To make the representation more complete and symmetric, 
it is necessary to somehow incorporate information about orientation of the plane 
with respect to the other crystal.

Such information is contained in a $4 \times 4$
interface matrix 
built of a $3 \times 3$ matrix representing the misorientation 
and boundary plane normals 
in reference frames of both crystals \cite{Morawiec_1998,Morawiec_2009,Morawiec_2004}. 
When generating input for atomistic simulations, the natural specification of  
boundaries relies on pairs of crystal orientations 
in a configuration in which the boundary plane has a fixed orientation in the
simulation box. In this context,  
Olmsted proposed to use pairs of orientation matrices \cite{Olmsted_2009_AB}. 
Olmsted's pairs of matrices were replaced by pairs of unit quaternions in \cite{Francis_2019}.
These extensive boundary representations 
are convenient in computations, but they are practically useless 
in interpersonal communication because of abundance of numbers involved
or intricate interpretation. 
Sutton and Balluffi \cite{Sutton_1995} proposed representing boundaries 
by misorientation and the linear combination $(\mathbf{n}+\mathbf{n}')/2$, 
where $\mathbf{n}$ and $\mathbf{n}'$ contain components of the unit vector
normal to the boundary plane in the reference frames of the two crystals, 
but the combination is not directly interpretable; 
one needs to refer to misorientation to get Miller indices of the boundary plane.

\subsection{Rotation matrices}

When dealing with crystal orientations, 
it is convenient to follow the Bunge's 
convention \cite{Bunge_1982} 
used in texture analysis and EBSD orientation mapping systems.
Orientations are identified with rotations relating  
Cartesian reference frames ascribed to the crystals and to the sample.
The rotation about axis along the unit vector 
$\mathbf{k}=[k^1 \ k^2 \ k^3 ]^T$ by the angle $\omega$ is represented by a 
$3 \times 3$ special orthogonal matrix
$R(\mathbf{k},\omega) = 
I  \cos \omega +\mathbf{k} \otimes \mathbf{k} \, (1- \cos \omega) + 
\mathbf{rot}(\mathbf{k}) \, \sin \omega$,
where the entries $ij$ of the identity matrix $I$,  $\mathbf{k} \otimes \mathbf{k}$
and 
$\mathbf{rot}(\mathbf{k})$
are $\delta_{ij}$,
 $k^i k^j$
and  $\sum_l \varepsilon_{ijl} k^l$, respectively.

In Bunge's convention, if the 
crystal orientation is represented by the matrix $g$
and the column array $\mathbf{n}_s$ contains vector coordinates in the sample reference frame, 
the coordinates in the crystal reference frame are $g \mathbf{n}_s$. 
The orientation matrix $g$ is usually parameterized using
Euler angles $\varphi_1$, $\phi$ and $\varphi_2$
$$
g = R(\varphi_1,\phi,\varphi_2) = 
R(\mathbf{e}_z, \varphi_2) R(\mathbf{e}_x,\phi) R(\mathbf{e}_z,\varphi_1) \ , 
$$
where $\mathbf{e}_x=[1\ 0\ 0]^T$ and $\mathbf{e}_z=[0\ 0\ 1]^T$.
Here, the notation is abused by using the same symbol $R$ for 
$R(\mathbf{k},\omega)$ and $R(\varphi_1,\phi,\varphi_2)$, but the meaning of 
the symbol can be easily inferred from the number and type of arguments.
Using $Q(\phi,\varphi_2)=R(0,\phi,\varphi_2)$, 
one also has $R(\varphi_1,\phi,\varphi_2) = 
Q(\phi,\varphi_2) R(\mathbf{e}_z,\varphi_1)$.

In calculations involving the interface matrices, it is convenient to use a $4 \times 4$ representation 
of rotations \cite{Morawiec_1998,Morawiec_2009,Morawiec_2004}.
The matrix corrsponding to $(\mathbf{k},\omega)$ is given by 
\bEq
\mathbf{R}(\mathbf{k},\omega)= 
\left[
\begin{array}{c|c}
	1   &   \mathbf{0}^T \\
	\hline 
	\mathbf{0}  &   R(\mathbf{k},\omega)   \\
\end{array} 
\right] \ , 
\label{eq:rot3to4}
\eEq
where $\mathbf{0}=[0 \, 0 \, 0]^T$.
If the arguments of $\mathbf{R}$ are three angles, the symbol denotes  
$
\mathbf{R}(\varphi_1,\phi,\varphi_2) = 
\mathbf{R}(\mathbf{e}_z,\varphi_2) 
\mathbf{R}(\mathbf{e}_x,\phi) \mathbf{R}(\mathbf{e}_z,\varphi_1)=
\mathbf{Q}(\phi,\varphi_2) \mathbf{R}(\mathbf{e}_z,\varphi_1)
$, where
$\mathbf{Q}(\phi,\varphi_2)=\mathbf{R}(0,\phi,\varphi_2)$.

\subsection{Interface matrix}

The misorientation between crystals $1$ and $2$, understood as a rotation relating the orientation $g_2$ of crystal $2$ to the orientation $g_1$ of crystal $1$, is represented by the matrix 
$M=g_1 g_2^T$. 
The boundary plane can be specified by the unit
vector $\mathbf{n}_1$ normal to the plane and directed (by
convention) towards the crystal $2$, with coordinates given in the Cartesian reference frame
of crystal~$1$.
The pair $(M, \mathbf{n}_1)$ unambiguously determines the macroscopic boundary geometry. 
The same boundary seen from the viewpoint of the other crystal
is determined by  $(M^T, \mathbf{n}_2)$, where the vector $\mathbf{n}_2=-M^T \mathbf{n}_1$ 
is directed towards the crystal $1$ and has coordinates given in the reference frame 
of crystal $2$. 
The interface matrix \cite{Morawiec_1998,Morawiec_2009,Morawiec_2004}
encompasses all $M$, $\mathbf{n}_1$ and $\mathbf{n}_2$: 
the matrix  $\widetilde{\mathbf{B}}$  representing the boundary between
crystals $1$ and $2$ is defined as
\bEq
\widetilde{\mathbf{B}}=
\mathbf{B}(M,\mathbf{n}_1 )= 
\left[
\begin{array}{c|c}
	  1   &   {\mathbf{n}_2}^T \\
	\hline 
	\mathbf{n}_1  &   M   \\
\end{array} 
\right] 
\ .
\label{eq:boundary_matrix}
\eEq
The boundary between crystals $2$ and $1$ is represented by $\widetilde{\mathbf{B}}^T$. 

Let the $4 \times 4$ matrices $\mathbf{R}_i$ ($i=1,2$)
be related to $3 \times 3$ matrices $R_i$ via (\ref{eq:rot3to4}). 
If crystal~$1$ is rotated by $R_1$, then $g_1$ and  $\mathbf{n}_1$
become  $R_1 g_1$ and $R_1 \mathbf{n}_1$, respectively. 
An analogous remark applies to crystal $2$.
Thus, rotations of crystal $1$ by $\mathbf{R}_1$
and crystal $2$ by $\mathbf{R}_2$
change the boundary represented by $\mathbf{B}(M,\mathbf{n}_1 )$ to the boundary 
represented by 
$\mathbf{B}(R_1 M R_2^T, R_1 \mathbf{n}_1 )=
\mathbf{R}_1 \mathbf{B}(M,\mathbf{n}_1 ) \mathbf{R}_2^T$.
This observation is convenient for taking into account crystal symmetries
\cite{Morawiec_1998,Morawiec_2009}.
Below, 
it is used 
for defining the new parameterization.

\section{The interface-plane parameterization}

\subsection{Relationships defining the parameterization}

Let
$\mathbf{B}_0$ denote the interface matrix $\mathbf{B}(I,\mathbf{e}_z)$.
Physically, $\mathbf{B}_0$ represents a configuration 
in which the first crystal occupies the region with $z<0$ 
and the second crystal with the same orientation occupies the region with $z>0$. 
An arbitrary interface matrix $\widetilde{\mathbf{B}}$
can be expressed as 
\bEq
\widetilde{\mathbf{B}}=\mathbf{R}_1 \mathbf{B}_0 \mathbf{R}_2^T \ . 
\label{eq:RBR}
\eEq
This relationship reflects the fact that an arbitrary boundary can be obtained by 
rotating the crystals forming the configuration represented by $\mathbf{B}_0$. 

The matrices $\mathbf{R}_i$ ($i=1,2$) are directly related to Olmsted's rotation matrices
defined in \cite{Olmsted_2009_AB}.
It must be emphasized that these are not orientation matrices 
obtained from an experiment. The experimental data must be rotated to 
the configuration with the boundary plane normal to 
the sample's $z$-direction. 
(The unit vector $\mathbf{n}_s$ normal 
to the boundary plane and directed 
outward from crystal $1$, with components in the sample reference frame,
is transformed to $\mathbf{e}_z$ by 
$R_0=R(\widehat{\mathbf{k}}_0,\omega_0)$ such that 
$\widehat{\mathbf{k}}_0$ is the normalized $\mathbf{k}_0=\mathbf{e}_z \times \mathbf{n}_s$
and $\omega_0=\arccos(\mathbf{e}_z \cdot \mathbf{n}_s)$
if $\mathbf{n}_s \neq \pm \mathbf{e}_z$, 
$\widehat{\mathbf{k}}_0$ is an arbitrary unit vector perpendicular to $\mathbf{e}_z$ 
and $\omega_0=\pi$ if $\mathbf{n}_s=-\mathbf{e}_z$,  and
$R_0=I$ if $\mathbf{n}_s=+\mathbf{e}_z$. 
One has $R_0 \mathbf{n}_s = \mathbf{e}_z$.
If experimentally obtained orientation of crystal $i$ is $g_i$ ($i=1,2$), 
then the matrices $\mathbf{R}_i$ used in (\ref{eq:RBR}) correspond to $R_i=g_i R_0^T$.)

\begin{figure}[t]
	\begin{picture}(300,280)(0,0)		
		\put(50, 0){\resizebox{12.0 cm}{!}{\includegraphics{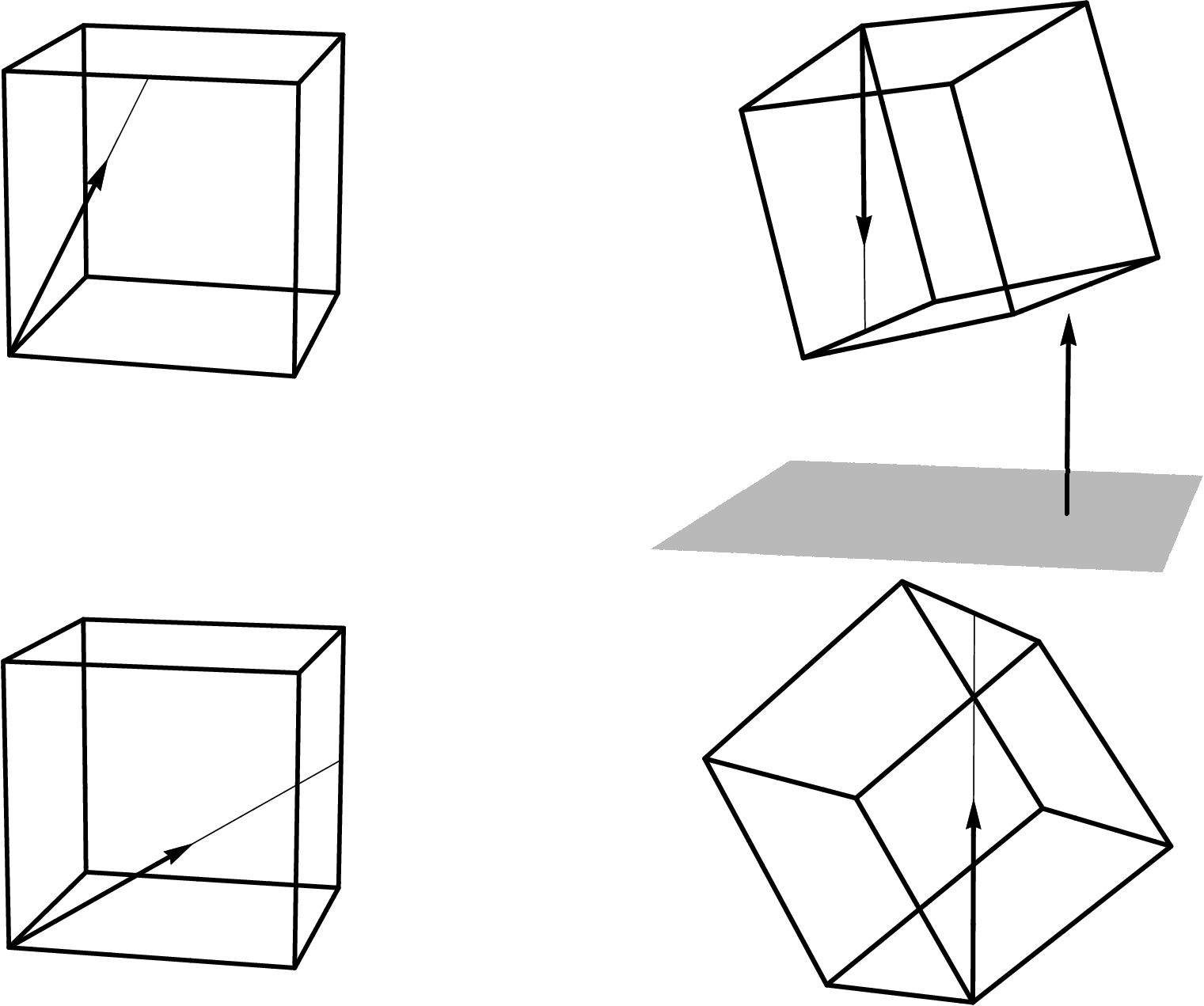}}}		
		\put(45,114){crystal 1}
		\put(45,170){crystal 2}		
		\put(180,225){$\mathbf{Q}(\theta_2,\psi_2)^T$}
		\put(175,220){\vector(1,0){55}}		
		\put(180,65){$\mathbf{Q}(\theta_1,\psi_1)$}
		\put(230,60){\vector(-1,0){55}}		
		\put(280,163){$R(\mathbf{e}_z,\chi)$}
		\put(276,174){\vector(0,-1){15}}		
		\put(356,170){$\mathbf{e}_z$}		
		\put(85,44){$\mathbf{n}_1$}
		\put(78,224){$\mathbf{n}_2$}		
		\put(205,142){boundary}
		\put(205,132){plane}		
		\put(52.5,15.8){\circle*{6}}
		\put(52.5,184){\circle*{6}}
		\put(325.7,1){\circle*{6}}
		\put(295,277){\circle*{6}}		
		\put(147,69.5){\circle*{5}}
		\put(92.2,262.8){\circle*{5}}
		\put(326,111){\circle*{5}}
		\put(295,191.5){\circle*{5}}		
		\put(352.5,140){\circle*{6}}
	\end{picture}
	\vskip 0.0cm
	\caption{Schematic illustration of the configuration used for defining the interface-plane parameters (in the case of cubic crystals). Coordinates of the vectors 
		$\mathbf{n}_i$ in the Cartesian reference frames linked to the crystals are related to $(\theta_i,\psi_i)$ via eq.(\ref{eq:EulerAnglesToVectors}) and $\chi$ is the angle of rotation about normal to the boundary plane $\mathbf{e}_z$. 
	}
	\label{Fig_1Dx}
\end{figure}
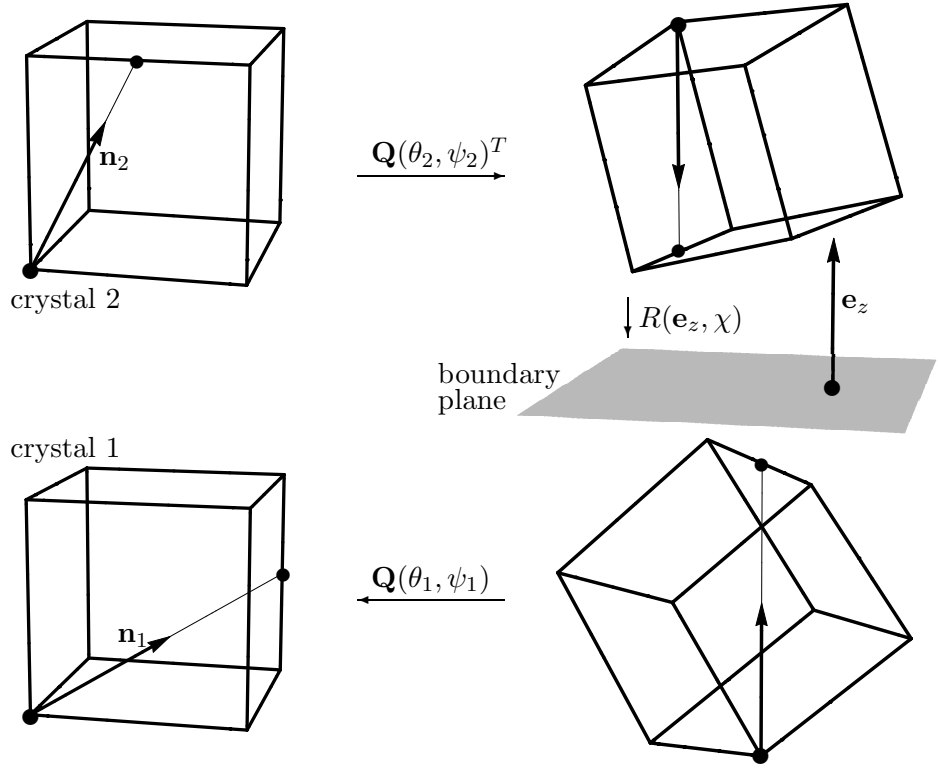

In the following, Euler angles parameterizing the rotations $\mathbf{R}_i$ will be needed.
The symbols  $\varphi_1$, $\phi$ and $\varphi_2$ are conventionally used 
for Euler angles of crystal orientations in the sample (i.e. for $g_i$). 
To avoid confusion, the Euler angles of $\mathbf{R}_i$ will be denoted by  
$\chi_i$, $\theta_i$ and $\psi_i$, i.e., $\mathbf{R}_i=\mathbf{R}(\chi_i,\theta_i,\psi_i)$.
In this notation, (\ref{eq:RBR}) has the form 
$$
\begin{array}{rcl}
\widetilde{\mathbf{B}}
 & = & 
\mathbf{R}(\chi_1,\theta_1,\psi_1)
\ \mathbf{B}_0 \ 
\mathbf{R}( \chi_2,\theta_2,\psi_2)^T
\\
 & = &
\mathbf{Q}(\theta_1,\psi_1) \mathbf{R}(\mathbf{e}_z,\chi_1)
\ \mathbf{B}( I,\mathbf{e}_z) \ 
\mathbf{R}(\mathbf{e}_z, \chi_2)^T
\mathbf{Q}(\theta_2,\psi_2)^T \ , \\
\end{array}
$$
and since 
$\mathbf{R}(\mathbf{e}_z,\chi_1)
\, \mathbf{B}( I,\mathbf{e}_z) \, 
\mathbf{R}(\mathbf{e}_z, \chi_2)^T
= \mathbf{B}\left( R( \mathbf{e}_z, \chi_1-\chi_2),\mathbf{e}_z \right)$,
one obtains
\bEq
\widetilde{\mathbf{B}} =
\mathbf{Q}(\theta_1,\psi_1)
\ \mathbf{B} \left( R( \mathbf{e}_z, \chi),\mathbf{e}_z \right) \, 
\mathbf{Q}(\theta_2,\psi_2)^T \ , 
\label{eq:basic_relationship}
\eEq
where 
$ \chi =\chi_1-\chi_2$. 
Thus, the interface matrix $\widetilde{\mathbf{B}}$ is expressed via five independent quantities:
$\theta_1,\psi_1, \,  \chi , \,  \theta_2,\psi_2$.
As (\ref{eq:basic_relationship}) can also be written in the form
$\widetilde{\mathbf{B}}=\mathbf{R}(\chi,\theta_1,\psi_1)
\, \mathbf{B}_0 \, 
\mathbf{R}(0,\theta_2,\psi_2)^T$,
$\chi$ can be interpreted as an Euler angle.
(See Fig.~\ref{Fig_1Dx}.)
Formulae for inverting eq.(\ref{eq:basic_relationship}), i.e.
for getting $\theta_1,\psi_1,  \chi ,   \theta_2,\psi_2$ 
from  $\widetilde{\mathbf{B}}$, are
listed in Appendix~A.

From eq.(\ref{eq:basic_relationship}) it follows that 
the first column of $\widetilde{\mathbf{B}}$
is determined by $\mathbf{Q}(\theta_1,\psi_1)$ and the first row 
is determined by $\mathbf{Q}(\theta_2,\psi_2)$. 
Hence, based on (\ref{eq:boundary_matrix}), the vectors $\mathbf{n}_1$ and $\mathbf{n}_2$ 
are expressible through 
$\theta_1,\psi_1$ and $\theta_2,\psi_2$, respectively. 
They are given by 
$\mathbf{n}_1 = +Q(\theta_1,\psi_1) \mathbf{e}_z$ and 
$\mathbf{n}_2 = -Q(\theta_2,\psi_2) \mathbf{e}_z$,  
or by
\bEq
\begin{array}{l}
\mathbf{n}_1 = + \ \Theta(\theta_1,\pi/2-\psi_1) \ \ \ \mbox{and} \ \ \ 
\mathbf{n}_2 = - \ \Theta(\theta_2,\pi/2-\psi_2) \ , 
\end{array}
\label{eq:EulerAnglesToVectors}
\eEq
where the function $\Theta$  relates the unit vector  $\mathbf{n}$
to its spherical coordinates $(\theta,\varphi)$ in the conventional way
\bEq
\mathbf{n}= \Theta(\theta,\varphi) = 
\left[\,
	\cos \varphi \sin \theta \ \, 
	\sin \varphi \sin \theta \ \, 
	\cos \theta \,
\right]^T \ .
\label{eq:vfromspherical}
\eEq
Instead of dealing with the quintuple
$(\theta_1,\psi_1, \, \chi, \, \theta_2,\psi_2)$,
it is more convenient to use the triplet 
$(\mathbf{n}_1 ,  \chi ,  \mathbf{n}_2)$.
This representation 
resembles the interface-plane scheme \cite{Wolf_1989}. 
However, unlike the interface-plane scheme,
it is a proper parameterization: 
every macroscopic boundary configuration is represented by a triplet, and
no
two macroscopically distinct boundaries correspond to the same triplet.
In the  case of zero misorientation angle, i.e., when $M=I$, 
one has $\mathbf{n}_1=-\mathbf{n}_2$ and $\chi=0$ 
or the triplet  
$(\mathbf{n} ,  0 , -\mathbf{n})$, where 
$\mathbf{n}$ is an arbitrary unit vector.
Such parameters do not represent true boundaries, 
but the considerations 
below  
are simpler if they are not excluded.

Although the parameter $\chi$ can be changed without affecting the crystal planes forming the boundary, it is not determined solely by the misorientation. In other words, 
boundaries with the same misorientation and different plane inclinations 
generally have different values of $\chi$. 
The configuration with 
$\chi=0$ corresponds to the misorientation 
represented by $R(-\psi_2,\theta_1-\theta_2, \psi_1)$.

Planes spanned by vectors of the direct crystal lattice 
can be represented by Miller indices. 
Let $\mathbf{n}$ be perpendicular to such a plane. 
To describe interfaces, 
Miller indices  
can be defined as integers $(h \, k \, l) = (h_1 \, h_2 \, h_3)$ such that
$h_j = \lambda \mathbf{n} \cdot \mathbf{a}_j$ 
or 
$\mathbf{n} = \sum_j h_j \mathbf{a}^j/\lambda$, 
where $\mathbf{a}_j$ ($\mathbf{a}^j$) denotes the 
$j$-th basis vector of the direct (reciprocal) crystal lattice,
and the real coefficient $\lambda$ is positive. 
If the plane of the boundary  $(\mathbf{n}_1 ,  \chi ,  \mathbf{n}_2)$ 
is expressible through Miller indices in 
each of the neighboring crystals,  
the boundary can be represented by 
$$
\left( \, (h \, k \, l)_1 , \, \chi , \, (h \, k \, l)_2 \, \right) \ , 
$$
where the Miller indices $(h \, k \, l)_i$ replaced the corresponding vector $\mathbf{n}_i$.

\subsection{Example}
\label{sec:Example}

It is worth illustrating the above considerations with an example.
Let the crystals have cubic $m\overline{3}m$ symmetry, 
and let axes of the crystal Cartesian coordinates systems be along four-fold symmetry axes.
The twin boundary common in fcc metals separates 
crystals misoriented by 
the half-turn about $\mathbf{k}=[1 \, 1 \, 1]^T/\sqrt{3}$, 
and the boundary plane is perpendicular to
$\mathbf{n}_1=[1 \, 1 \, 1]^T/\sqrt{3}$. The corresponding 
interface matrix has the form 
$$
\mathbf{B}\left( R(\mathbf{k},\pi), \mathbf{n}_1 \right)=
\frac{1}{3} \left[
\begin{array}{r|rrr}
	1 & -\sqrt{3} & -\sqrt{3} & -\sqrt{3} \\
	\hline
	\sqrt{3} & -1 &  2 &  2 \\
	\sqrt{3} &  2 & -1 &  2 \\
	\sqrt{3} &  2 &  2 & -1 \\
\end{array}
\right] \ . 
$$
Hence, one has
$$
(\theta_1,\psi_1, \,  \chi , \,  \theta_2,\psi_2) =
\left(
\arctan \sqrt{2},\pi/4, \,
\pi, \, 
\arctan \sqrt{2}, \pi/4
\right) \ , 
$$
or easier to interpret
$
(\mathbf{n}_1 ,  \chi ,   \mathbf{n}_2) = 
\left([1 \, 1 \, 1]^T/\sqrt{3} \, , \  \pi \, , \ -[1 \, 1 \, 1]^T/\sqrt{3} \right)
$.
Replacing the normalized vectors $\mathbf{n}_i$ with
Miller indices gives
$$(\mathbf{n}_1 ,  \chi ,   \mathbf{n}_2) \sim 
\left( \, (1 \, 1 \, 1)  , \  \pi  , \ 
(\overline{1} \, \overline{1} \, \overline{1}) \, \right) \ . 
$$
Due to crystal symmetry, 
this is one of several equivalent representations of the twin boundary.

\subsection{Equivalences due to symmetries}

Because of symmetries, different sets of parameters stand for the same
macroscopic boundary configuration.
The solution to the problem of equivalent descriptions of boundaries between centrosymmetric 
crystals is given in \cite{Morawiec_1998}; see also \cite{Morawiec_2009}.
The equivalences listed there can be easily rewritten in the new formalism.
The assumption that the crystal point group contains inversion 
means that 
\bEq
(\mathbf{n}_1 ,  \chi , \mathbf{n}_2) \simeq
(-\mathbf{n}_1 , - \chi , -\mathbf{n}_2) \ , 
\label{eq:equiv_inversion}
\eEq
where $\simeq$ stands for equivalence. 
With $C_1$ and $C_2$ denoting matrices representing proper rotations 
in the point group of the crystal, 
one has
\bEq
(\mathbf{n}_1 ,  \chi , \mathbf{n}_2) \simeq 
(C_1 \mathbf{n}_1 ,  \chi' , C_2 \mathbf{n}_2) \ ,
\label{eq:equiv_CBC}
\eEq
where $\chi'$, in general, differs from $\chi$.
The most convenient way to determine 
$\chi'$ is to transform $(\mathbf{n}_1 ,  \chi , \mathbf{n}_2)$ to 
$\widetilde{\mathbf{B}}$,
get  $\widetilde{\mathbf{B}}'=\mathbf{C}_1 \widetilde{\mathbf{B}}\mathbf{C}_2^T \simeq  \widetilde{\mathbf{B}}$,
and transform $\widetilde{\mathbf{B}}'$
back to $(C_1 \mathbf{n}_1 ,  \chi' , C_2 \mathbf{n}_2)$;
$\mathbf{C}_i$ is the $4 \times 4$ matrix 
related to $C_i$ via eq.(\ref{eq:rot3to4}).

The equivalences
(\ref{eq:equiv_inversion}) and (\ref{eq:equiv_CBC}) 
arise from the symmetry of the crystal and are therefore inevitable. 
Depending on circumstances, one may or may not allow for grain exchange symmetry.
Allowing this symmetry means that the boundary between 
grains $1$ and $2$ is identified with the boundary between grains $2$ and $1$.
In terms of  the interface-plane parameters, the grain exchange symmetry has the form 
\bEq
(\mathbf{n}_1 ,  \chi , \mathbf{n}_2) \simeq 
(\mathbf{n}_2 ,  \chi , \mathbf{n}_1) \ .
\label{eq:equiv_grain_exchange}
\eEq
Like 
(\ref{eq:equiv_inversion}) and (\ref{eq:equiv_CBC}),
also this relationship
follows directly from the corresponding 
equivalence expressed via interface matrices 
in \cite{Morawiec_1998}.

It is worth stating explicitly that, in general, 
the triplet $(\mathbf{n}_1,  \chi, \mathbf{n}_2)$ is not equivalent 
to $(\mathbf{n}_1,  \chi', -\mathbf{n}_2)$.
More precisely, if none of the vectors $\mathbf{n}_i$ is transformed to  
$-\mathbf{n}_i$ by a symmetry operation  being a proper rotation, then 
there are no such $\chi$ and  $\chi'$ that 
$(\mathbf{n}_1,  \chi, \mathbf{n}_2) \simeq  (\mathbf{n}_1,  \chi', -\mathbf{n}_2)$.
For example, if crystals have $m\overline{3}m$ symmetry, 
the representation
$\left( (1\, 2 \, 3), \chi,  (3\, 5 \, 7)  \right)$ 
is never equivalent to 
$\left( (1\, 2 \, 3), \chi',  (\overline{3}\, \overline{5} \, \overline{7})  \right)$.
Similarly,  
$\left( (1\, 2 \, 3), \chi,  (3\, 5 \, 7)  \right)$ 
is never equivalent to $\left( (1\, 2 \, 3), \chi',  (7 \, 5 \, 3)  \right)$.
In view of the above, it is clear that if boundary planes 
are represented by Miller indices, 
a distinction must be made between $(h\, k\, l)$ and 
$(\overline{h}\, \overline{k} \, \overline{l})$, and between different sequences of indices.
Awareness of this fact is not widespread
among those working on cubic materials.
It is symptomatic that the works focused on pairs of  
crystal planes forming a boundary (e.g., \cite{Randle_1997,Randle_2006,Wang_2017,Wang_2018,Wang_2019}) ignore the necessity of making such distinctions.
This necessity is somewhat hidden 
when boundaries 
are represented by misorientation/plane pairs, but is
evident when the interface-plane parameterization is used.

Based on  
(\ref{eq:equiv_inversion}) and (\ref{eq:equiv_CBC}), 
it can be verified that 
the representations equivalent 
to the example twin boundary $(\mathbf{n}_1 ,  \chi ,   \mathbf{n}_2) \sim 
\left( (1 \, 1 \, 1)  , \  \pi  , \ 
(\overline{1} \, \overline{1} \, \overline{1}) \right)$ 
described in section \ref{sec:Example}
have the forms
\bEq
	\left( (h \, k \, l), \ j \pi/3  , \  (h' \, k' \, l') \right) \ , 
\label{eq:equiv_to_example}
\eEq
where the indices $h,k,l,h',k',l'$  
take the values of $\pm 1$, $j=0,1,\ldots,5$, 
and the condition 
$ l  l' = 1 - 2 \, \mbox{mod}(j,2)$ 
is satisfied, 
i.e., $l  l' = +1$ when $j$ is even, and $l  l' = -1$ when $j$ is odd.
For instance, 
with all Miller indices equal to $+1$ and $j=0$,
one has
\bEq
\left( (1 \, 1 \, 1)  , \  \pi  , \ 
(\overline{1} \, \overline{1} \, \overline{1}) \right) \simeq
 \left( (1 \, 1 \, 1)  , \  0  , \ (1 \, 1 \, 1) \right) \ .
\label{eq:equiv_twin_twist}
\eEq
The number of distinct representations (\ref{eq:equiv_to_example}) 
is $2^6 \times  3 = 192$.
The number of proper rotations 
in the point group $m\overline{3}m$ is $N_p=24$, and   
the number of equivalent representations of a general boundary
is $2 \times N_p^2= 1152$. 
Thus,  the case of twin boundary
is degenerate, and the degree of degeneracy is 6.

\subsection{Fundamental region}

Since  each of the vectors $\mathbf{n}_1$ and $\mathbf{n}_2$ covers the unit sphere $S^2$ 
and $\chi$ covers the circle $S^1$, 
the domain $D$ of the triplets 
$(\mathbf{n}_1 , \chi , \mathbf{n}_2)$ is the Cartesian product $S^2 \times S^1 \times S^2$.
Due to symmetries,
physical boundaries correspond to classes of equivalent parameters. 
These classes are analogous to classes of equivalent crystallographic directions or 
equivalent orientations typically dealt with in texture analysis.
In such cases, 
computations rely on representatives of equivalence classes.
It is convenient to specify the part of the domain $D$ 
in which these representatives are gathered \cite{Morawiec_2009}.
Such part is known as the fundamental region.
It is a closed subset of $D$
such that the images of the subset due to the symmetries 
cover $D$, and the interiors of the
images have no common points. 
Determination of fundamental regions for
$(\mathbf{n}_1,\chi,\mathbf{n}_2)$-parameterized boundaries 
relies on fundamental regions for crystallographic directions.

For illustration, inequalities determining a fundamental region for 
$m\overline{3}m$ crystal symmetry will be described. 
In this case, 
the fundamental region for crystallographic directions is known as 
the standard stereographic triangle (SST).
It can be defined as a set of 
directions $\mathbf{n}_i = [n_i^1 \, n_i^2 \, n_i^3]^T$ ($i=1,2$) satisfying the conditions 
$n_i^3 \geq n_i^2 \geq n_i^1 \geq 0$. 
Proper rotations  of $m\overline{3}m$ constitute the point group $432$, 
for which the fundamental region of directions 
consists of the double standard stereographic triangle (DSST); 
the latter can be defined by the inequalities  
$n_i^3 \geq n_i^k \geq 0$, where $k=1,2$. 

\begin{figure}[t]
	\begin{picture}(300,200)(0,0)
		\put(-40,0){\resizebox{7.0 cm}{!}{\includegraphics{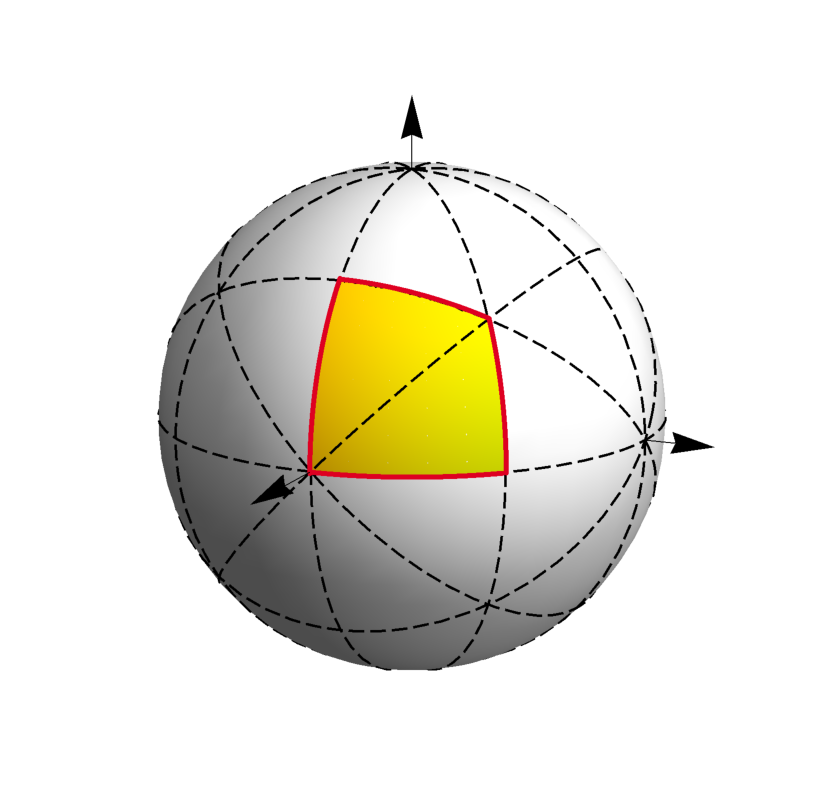}}}
		\put(260,0){\resizebox{7.0 cm}{!}{\includegraphics{Fig_2.eps}}}		
		\put(157,30){\resizebox{4.0 cm}{!}{\includegraphics{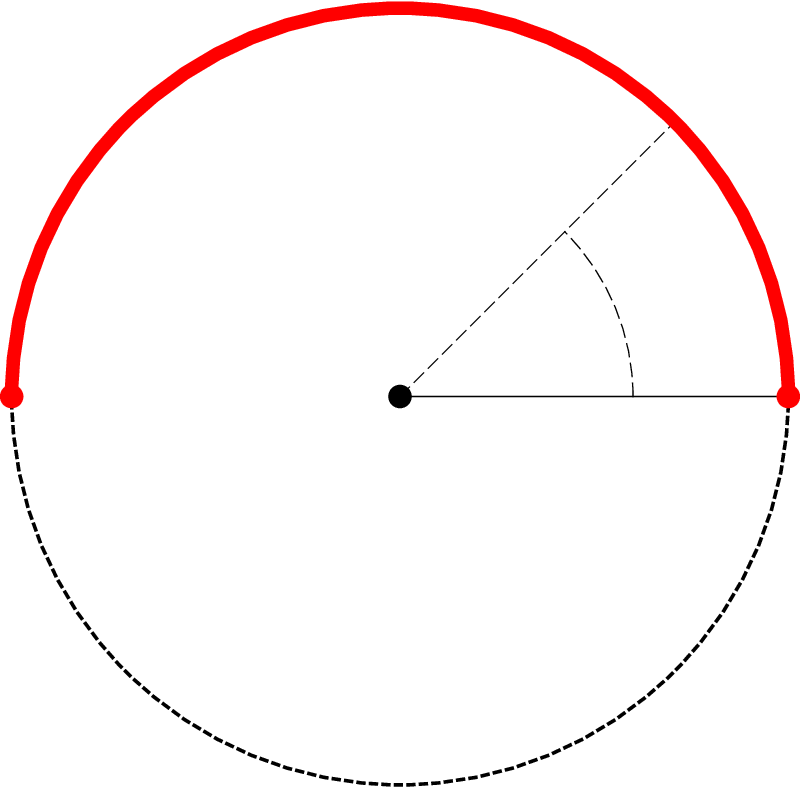}}}
		\put(54,190){\Large $\mathbf{n}_1$}
		\put(207,190){\Large $\chi$}
		\put(353,190){\Large $\mathbf{n}_2$}
		\put(47,5){\Large DSST}
		\put(135,5){\huge $\times$}
		\put(207,5){\Large $S^1_+$}
		\put(277,5){\huge $\times$}
		\put(343,5){\Large DSST}
		\put(122,75){$\mathbf{e}_x$}
		\put(63,160){$\mathbf{e}_y$}
		\put(422,75){$\mathbf{e}_x$}
		\put(363,160){$\mathbf{e}_y$}
		\put(135,81){\huge $\times$}
		\put(277,81){\huge $\times$}
	\end{picture}
	\caption{Schematic illustration of the fundamental region (\ref{eq:fundamtal_region_1})
for $(\mathbf{n}_1,\chi,\mathbf{n}_2)$-parameterized boundaries between 
crystals with $m\overline{3}m$ symmetry. 
This region concerns the mandatory equivalences 
(\ref{eq:equiv_inversion}) and (\ref{eq:equiv_CBC}). 
The areas marked on the spheres are double standard stereographic triangles
defined by $n_i^3 \geq n_i^k \geq 0$, where $i=1,2$ and $k=1,2$. 
}
	\label{Fig_FR}
\end{figure}

If only the equivalences (\ref{eq:equiv_CBC}) 
are used, the fundamental region for
$(\mathbf{n}_1,\chi,\mathbf{n}_2)$
is $\mbox{DSST} \times S^1 \times \mbox{DSST}$, and  
eq.(\ref{eq:equiv_inversion})  reduces it further  
to 
\bEq
\mbox{DSST} \times S^1_+ \times \mbox{DSST} \ , 
\label{eq:fundamtal_region_1}
\eEq
where $S^1_+$ denotes the semicircle defined by 
$\chi$ ranging from $0$ to $\pi$; see Fig.~\ref{Fig_FR}.

The choice of the fundamental region is not unique. 
The region (\ref{eq:fundamtal_region_1}) can be replaced by 
$\mbox{SST} \times S^1 \times \mbox{DSST}$ or $\mbox{DSST} \times S^1 \times \mbox{SST}$. 
Yet another option is to use 
$\mbox{DSST} \times S^1 \times \mbox{DSST}$ with the condition 
$n_1^1-n_1^2 \geq n_2^1-n_2^2$, but it must be noted that the form of 
this condition depends on the inequalities defining DSST.\footnote{For instance, 
if DSST is defined by $n_i^3 \geq n_i^2 \geq | n_i^1 |$, one can use the condition  
$n_1^1 \geq n_2^1$.}
The latter option 
is mentioned because a similar scheme 
is necessary to account for grain exchange symmetry.

If besides  (\ref{eq:equiv_inversion}) and (\ref{eq:equiv_CBC})
also the grain exchange symmetry  (\ref{eq:equiv_grain_exchange}) holds, the 
fundamental region for interface-plane parameters is 
$\mbox{DSST} \times S^1 \times \mbox{DSST}$ with the conditions 
$n_1^1-n_1^2 \geq n_2^1-n_2^2$ and $n_1^1+n_1^2 \geq n_2^1+n_2^2$.\footnote{If DSST 
is defined by $n_i^3 \geq n_i^2 \geq | n_i^1 |$, one can use the conditions 
$n_1^1 \geq n_2^1$ and $n_1^2 \geq n_2^2$.}

These expressions for the fundamental regions were confirmed by numerical tests.

\subsection{Characteristic boundaries}

Characteristic boundaries
have special arrangements of boundary planes and misorientation axes.
Usually, specifications of such boundaries concern low--index cases, 
but formal definitions ignore magnitudes of the indices.
In particular, a boundary having a representation with the misorientation axis 
perpendicular (parallel) to the boundary plane is a twist (tilt) boundary.
A boundary is symmetric when the boundary plane is a mirror plane between crystal structures.

The  interface-plane parameters of twist and symmetric boundaries are simple. 
A symmetric boundary is represented by 
$$
(\mathbf{n} , \pi, -\mathbf{n}) \ , 
$$
and every triplet having this form represents a symmetric boundary. 
Such a boundary is determined by $\mathbf{n}$, 
i.e.,  by two independent quantities \cite{Morawiec_2012}. 
The triplet
\bEq
(\mathbf{n} ,  \chi , - \mathbf{n})
\label{eq:twist_b}  
\eEq
represents a twist boundary.
The misorientation angle of a boundary represented by $(\mathbf{n} ,  \chi ,  -\mathbf{n})$ 
is $\arccos( \cos \chi)$, and its
misorientation axis is $\mbox{sign}(\sin \chi) \, \mathbf{n}$
if $\chi$ differs from $0$ and $\pi$; the axis is $\pm \mathbf{n}$ if $\chi=\pi$,
and is arbitrary if $\chi=0$.
The misorientation angle of a boundary represented by
\bEq
(\mathbf{n} ,  \chi , + \mathbf{n})  
\label{eq:180_tilt_b} 
\eEq
is $\pi$, 
and its misorientation axis is perpendicular to $\mathbf{n}$, i.e.,  
(\ref{eq:180_tilt_b}) represents a $180^{\circ}$-tilt boundary.
Every $180^{\circ}$-tilt boundary has a representation of this form.
Boundaries represented by triplets of types (\ref{eq:twist_b}) and (\ref{eq:180_tilt_b})
are determined by three independent quantities \cite{Morawiec_2012}.

Representations of general tilt boundaries 
are more complex;
see, Appendix~B. 
For a boundary to be called a tilt boundary, 
it must satisfy one scalar condition, 
therefore tilt boundaries form a four-dimensional subspace of all boundaries.

Clearly, when characterizing  experimental boundary data, 
it is not enough to look at a single set of parameters, 
but one also needs to consider symmetrically equivalent configurations. 
E.g., in order to test whether the boundary with parameters 
$(\mathbf{n}_1,  \chi ,  \mathbf{n}_2)$
is a twist boundary, one needs to check not only if 
$\mathbf{n}_1 \approx - \mathbf{n}_2$, but also whether
there are $C_1$  and $C_2$ 
such that  
$C_1 \mathbf{n}_1 \approx - C_2 \mathbf{n}_2$. 
Taking this into account, it is easy to see that a given  
boundary can have several attributes, e.g., 
it can be a twist as well as a tilt boundary \cite{Morawiec_2009b}. 
A prime 
example of this case is the twin boundary described in section \ref{sec:Example}; 
its equivalent representations (\ref{eq:equiv_twin_twist}) 
have the forms (\ref{eq:twist_b}) and  (\ref{eq:180_tilt_b}).

\subsection{Uniform boundary distribution}

One of the basic characteristics of boundary networks 
in polycrystalline materials is the 
frequency of occurrence of particular boundary types \cite{saylor2002distribution}. 
For gauging the frequencies, 
a model representing the uniform distribution of boundaries is needed. 
The model is directly linked to the method of 
generating uniformly distributed (`random') boundaries.
   
The simplest and convenient model assumes that a random boundary is obtained by 
using a random misorientation and a random 
boundary plane \cite{Morawiec_2009}.  
The question is how this model translates to the 
parameterization by $(\mathbf{n}_1,\chi, \mathbf{n}_2)$? 
Based on properties of 
proper rotations (representing misorientations), 
the model corresponds to uniform distributions of 
$\mathbf{n}_1$ and  $\mathbf{n}_2$  on the sphere $S^2$ and 
uniform distribution of $\chi$ on the circle $S^1$. 
In other words,  if the triplet $(\mathbf{n}_1,\chi, \mathbf{n}_2)$
representing a random boundary is needed, 
it is enough to generate random directions $\mathbf{n}_1$ and $\mathbf{n}_2$, 
and a random angle $\chi$. 
Technically, with $\mbox{rnd}{(a,b)}$ denoting a random real number between $a$ and $b$,
the random directions are $\mathbf{n}_i=\Theta(\theta_i, \psi_i)$, where 
$\theta_i=\arccos \mbox{rnd}{(-1,1)} $ and $\psi_i=\mbox{rnd}{(0,2 \pi)}$, 
and $\chi=\mbox{rnd}{(0,2 \pi)}$.

The above model of uniformity is related to the differential volume element $\mbox{d}V$, 
which is a tool for integrating functions
over macroscopic boundary parameters. 
It is given by 
$$
\mbox{d}V = 
\mbox{d}\mathbf{n}_1 \,  \mbox{d}\mathbf{n}_2 \, \mbox{d} \chi/(2\pi) =
\sin \theta_1 \sin \theta_2 \   
\mbox{d}\theta_1  \, \mbox{d} \psi_1  \, \mbox{d} \chi  \, 
\mbox{d} \theta_2  \, \mbox{d} \psi_2/(32 \pi^3) \ . 
$$
This expression follows directly from  
the formula for area element on unit sphere in spherical coordinates.

\section{Grain boundary energy versus interface-plane parameters}

Parameterizations of grain boundaries allow for estimating 
the dependence of boundary properties on boundary types.
In this way, grain boundary  populations, energies and curvatures
as functions of grain misorientation and boundary plane have been evaluated;
see, e.g., \cite{saylor2002distribution,saylor2003relative,Zhong2017}.
Using different parameterizations allows for 
looking at these functions from different perspectives.

As an example, the dependence of the boundary free energy 
on the interface-plane parameters 
is considered.
The celebrated Bulatov-Reed-Kumar (BRK) model  \cite{bulatov2014grain} is used.
It approximates atomic-scale simulation data taken from \cite{olmsted2009survey}.
The case of Ni is inspected.
The model is denoted here by $\Gamma$, 
and $\Gamma|_{(\mathbf{n}_1,\mathbf{n}_2)}$ denotes the restriction  
of $\Gamma$
for fixed $\mathbf{n}_1$ and $\mathbf{n}_2$ and variable $\chi$.
Clearly, the function $\Gamma$ takes the same values for equivalent boundary representations, 
and this holds not only for the equivalences 
(\ref{eq:equiv_inversion}) and (\ref{eq:equiv_CBC}) 
but also for~(\ref{eq:equiv_grain_exchange}).

\begin{table}[htbp]
	\centering
	\begin{tabular}{cc|ccc}
		$\mathbf{n}_1$ & $\mathbf{n}_2$ 	& min &
		$\overline{\Gamma}|_{(\mathbf{n}_1,\mathbf{n}_2)}$  	& max \\ 
		\hline
		$(1 \, 0 \, 0)$ & $(1 \, 0 \, 0)$ &      & 0.80 & 0.99 \\ 		
		$(1 \, 0 \, 0)$ & $(1 \, 1 \, 0)$ & 0.90 & 1.24 & 1.37 \\ 		
		$(1 \, 0 \, 0)$ & $(1 \, 1 \, 1)$ & 0.76 & 0.96 & 1.02 \\ 		
		$(1 \, 0 \, 0)$ & $(2 \, 1 \, 0)$ & 1.23 & 1.27 & 1.30 \\ 		
		$(1 \, 0 \, 0)$ & $(2 \, 1 \, 1)$ & 1.00 & 1.14 & 1.17 \\ 		
		$(1 \, 0 \, 0)$ & $(2 \, 2 \, 1)$ & 0.78 & 1.10 & 1.27 \\ 		
		$(1 \, 1 \, 0)$ & $(1 \, 1 \, 0)$ &      & 1.07 & 1.31 \\ 		
		$(1 \, 1 \, 0)$ & $(1 \, 1 \, 1)$ & 1.17 & 1.26 & 1.28 \\  		
		$(1 \, 1 \, 0)$ & $(2 \, 1 \, 0)$ & 1.03 & 1.11 & 1.17 \\ 		
		$(1 \, 1 \, 0)$ & $(2 \, 1 \, 1)$ & 0.87 & 1.20 & 1.29 \\ 		
		$(1 \, 1 \, 0)$ & $(2 \, 2 \, 1)$ & 1.04 & 1.19 & 1.30 \\ 		
		$(1 \, 1 \, 1)$ & $(1 \, 1 \, 1)$ &      & 0.41 & 0.50 \\ 		
		$(1 \, 1 \, 1)$ & $(2 \, 1 \, 0)$ & 1.17 & 1.24 & 1.31 \\ 		
		$(1 \, 1 \, 1)$ & $(2 \, 1 \, 1)$ & 0.90 & 1.14 & 1.27 \\ 		
		$(1 \, 1 \, 1)$ & $(2 \, 2 \, 1)$ & 0.89 & 1.08 & 1.23  \\ 		
		$(2 \, 1 \, 0)$ & $(2 \, 1 \, 0)$ &      & 1.11 & 1.35 \\ 		
		$(2 \, 1 \, 0)$ & $(2 \, 1 \, 1)$ & 1.05 & 1.19 & 1.30 \\ 		
		$(2 \, 1 \, 0)$ & $(2 \, 2 \, 1)$ & 0.94 & 1.21 & 1.33 \\ 		
		$(2 \, 1 \, 1)$ & $(2 \, 1 \, 1)$ &      & 1.12 & 1.35 \\ 		
		$(2 \, 1 \, 1)$ & $(2 \, 2 \, 1)$ & 0.59 & 1.13 & 1.29 \\ 		
		$(2 \, 2 \, 1)$ & $(2 \, 2 \, 1)$ &      & 1.09 & 1.30 \\ 		
		$(1 \, 2 \, 3)$ & $(3 \, 5 \, 7)$ & 0.65 & 1.13 & 1.30 \\  		
		$(1 \, 2 \, 3)$ & $(\overline{3} \, \overline{5} \, \overline{7})$ & 0.33  & 1.16  
		&  1.30 \\  
	\end{tabular}
	\caption{Energy of boundaries in Ni based on the BRK model for 
		fixed $\mathbf{n}_1$ and $\mathbf{n}_2$. 
		$\overline{\Gamma}|_{(\mathbf{n}_1,\mathbf{n}_2)}$ 
		is the energy averaged over $\chi$, 
		and min and max are the maximum and minimum values of $\Gamma|_{(\mathbf{n}_1,\mathbf{n}_2)}$, respectively. 
		The min values in empty cells are zero; 
		they correspond to the `no boundary' case.
		The units are $\mbox{J/m}^2$. For brevity, 
		the vectors $\mathbf{n}_i$ are replaced by 
		Miller indices of the boundary planes.
	}
	\label{table:1}
\end{table}

With $\Gamma$ expressed using the interface-plane parameterization, 
one can easily get the restriction  $\Gamma|_{(\mathbf{n}_1,\mathbf{n}_2)}$  
and look at the energies of boundaries with the planes
characterized by $\mathbf{n}_1$ and $\mathbf{n}_2$.
Tab.~\ref{table:1} contains minimal, maximal and average values of 
$\Gamma|_{(\mathbf{n}_1,\mathbf{n}_2)}$ 
for a number of pairs $(\mathbf{n}_1,\mathbf{n}_2)$. 
It is noteworthy that
the energy of boundaries formed by two $(1\, 1\, 1)$ planes 
not only has a low average value, 
but also the maximum energy reached in this case is low.
On the other hand, maximum energies of 
boundaries formed by two $(1\, 1\, 0)$ planes 
or two $(2\, 1\, 1)$ planes are high. 
The boundaries formed by $(1\, 1\, 0)$ and $(1\, 1\, 1)$
have high minimum and average energies.
The last two rows of Tab.~\ref{table:1} for high-index boundary planes are to illustrate 
the lack of equivalence between the pairs $(\mathbf{n}_1, \mathbf{n}_2)$
and $(\mathbf{n}_1, -\mathbf{n}_2)$.

\begin{figure}
	\begin{picture}(300,460)(0,0)
		\put(60,220){\resizebox{10.0 cm}{!}{\includegraphics{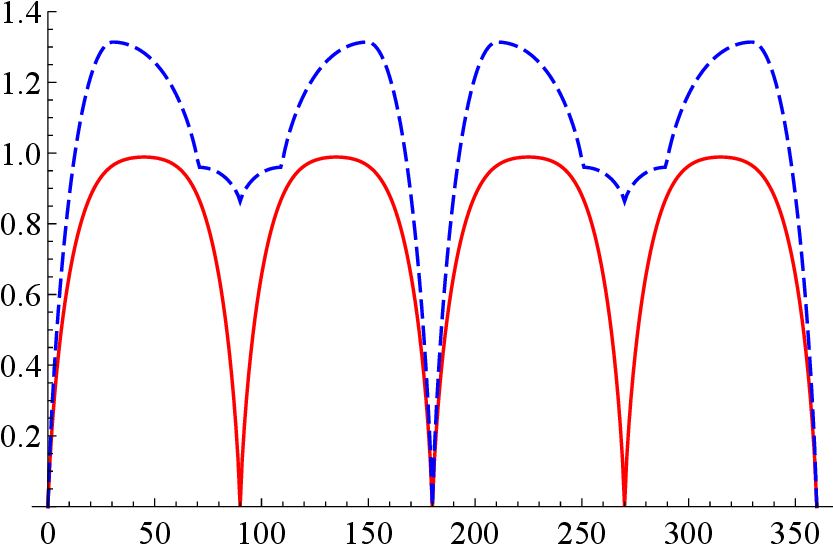}}}		
		\put(60,0){\resizebox{10.0 cm}{!}{\includegraphics{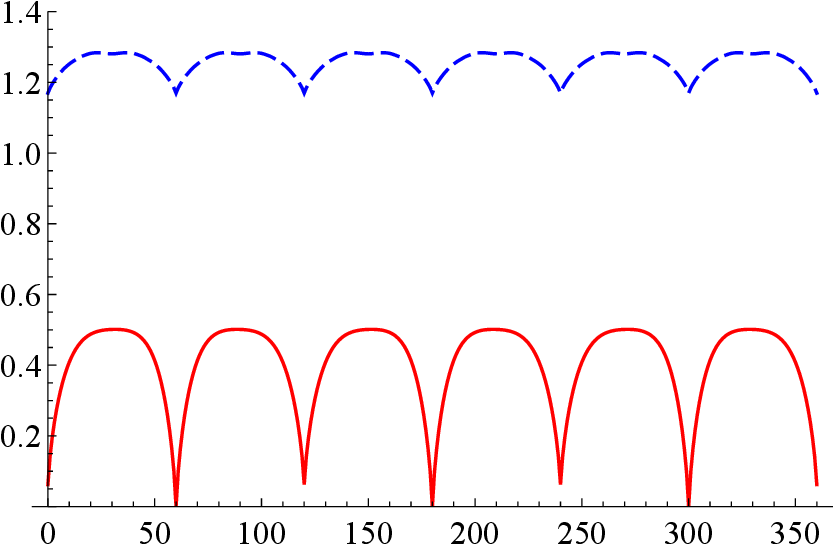}}}		
		\put(5,395){\textit{a}}
		\put(5,175){\textit{b}}		
		\put(346,222){$\chi$ [$\null^{\circ}$]}
		\put(20,412){$\Gamma|_{(\mathbf{n}_1,\mathbf{n}_2)}$ [$\mbox{J/m}^2]$}		
		\put(145,260){\small \textcolor{red}{$((100),(100))$}}
		\put(168,397){\textcolor{blue}{\large $((1\,1\,0),(1\,1\,0))$}}		
		\put(104.7,400){\color{blue}\line(0,-1){8}}
		\put(97,401){\textcolor{blue}{\small $\Sigma9$}}
		\put(113.17,394){\color{blue}\line(0,-1){8}}
		\put(112.5,394.5){\textcolor{blue}{\small $\Sigma11$}}
		\put(207.5,355){\color{blue}\line(0,-1){40}}
		\put(202.5,356){\small \textcolor{blue}{$\Sigma1$}}
		\put(273.3,395){\color{blue}\line(0,-1){50}}		
		\put(269,396){\textcolor{blue}{\footnotesize $R([1\overline{1}\sqrt{2}]^T/2,\pi)$}}
		\put(273.3,315){\color{red}\line(0,-1){30}}
		\put(268,316){\small \textcolor{red}{$\Sigma1$}}
		\put(142.0,315){\color{red}\line(0,-1){30}}
		\put(136.7,316){\small \textcolor{red}{$\Sigma1$}}
		\put(103.3,350){\color{red}\line(0,-1){20}}
		\put(98.0,322){\small \textcolor{red}{$\Sigma5$}}		
		\put(346,2){$\chi$ [$\null^{\circ}$]}
		\put(20,194){$\Gamma|_{(\mathbf{n}_1,\mathbf{n}_2)}$ [$\mbox{J/m}^2]$}		
		\put(230,79){\textcolor{red}{\large $((1\,1\,1),(1\,1\,1))$}}
		\put(230,142){\textcolor{blue}{\large $((1\,1\,0), (1\,1\,1))$}}
		\put(180,115){\textcolor{blue}{\footnotesize  
		$R([\overline{1}\, 1 \, 0]^T/\sqrt{2},\arccos(-\sqrt{2/3}))$}}
		\put(208,126){\color{blue}\line(0,1){25}}
		\put(114,79){\textcolor{red}{\small $\Sigma1$}}
		\put(120,77){\color{red}\line(0,-1){25}}
		\put(157,79){\textcolor{red}{\small $\Sigma3$}}
		\put(163.6,77){\color{red}\line(0,-1){25}}
		\put(143,92){\textcolor{red}{\small $\Sigma7$}}
		\put(148.5,90){\color{red}\line(0,-1){15.5}}
	\end{picture}
	\captionsetup{labelformat=empty}
	\vskip 0.0cm
	\caption{\\ \textit{Continued on next page \ \ \ } 
	}
\end{figure}

\addtocounter{figure}{-1}

\begin{figure}
	\begin{picture}(300,410)(0,0)		
		\put(60,220){\resizebox{10.0 cm}{!}{\includegraphics{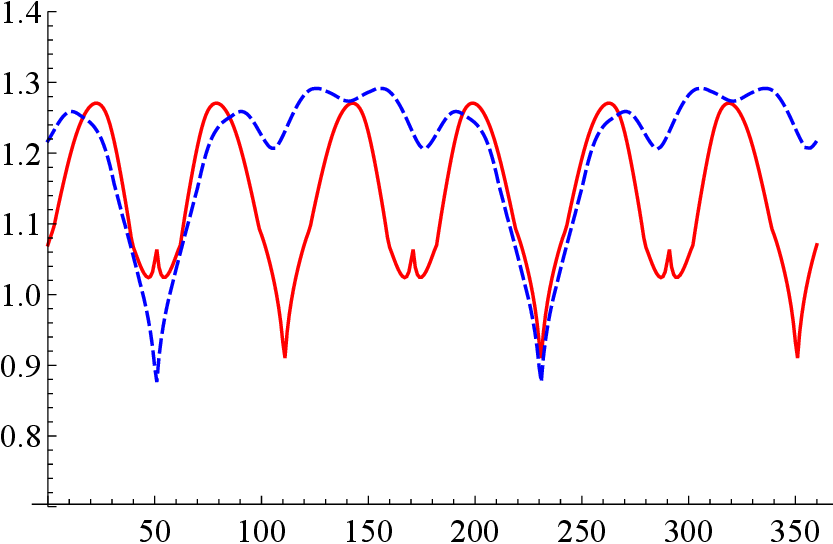}}}
		\put(280,340){\resizebox{3.6 cm}{!}{\includegraphics{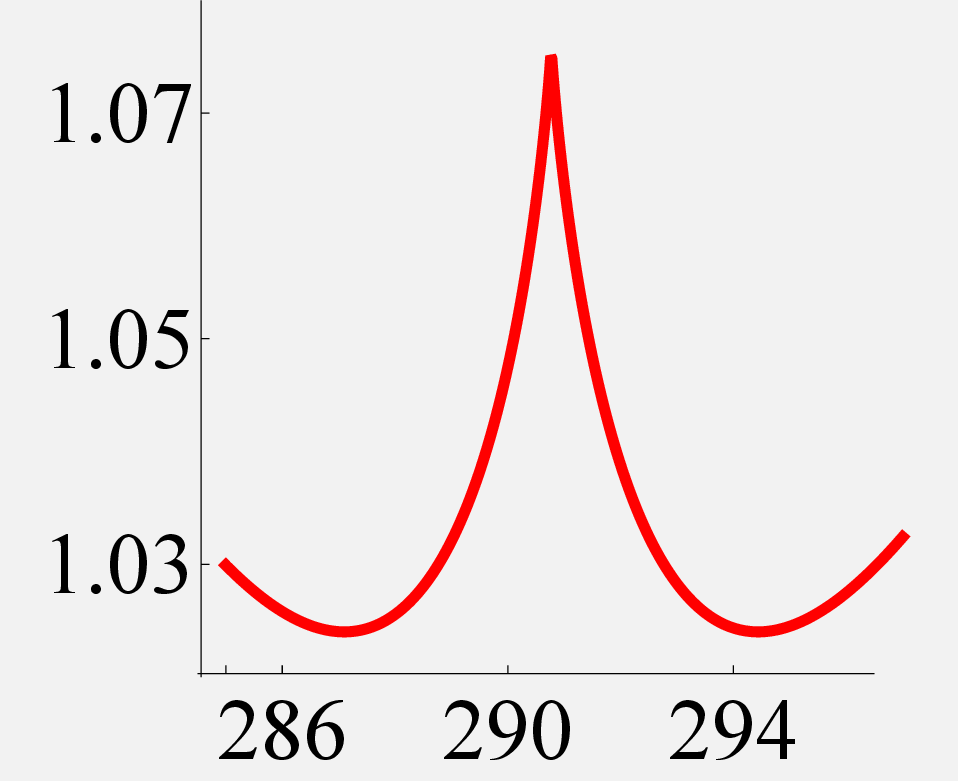}}}
		\put(60,0){\resizebox{10.0 cm}{!}{\includegraphics{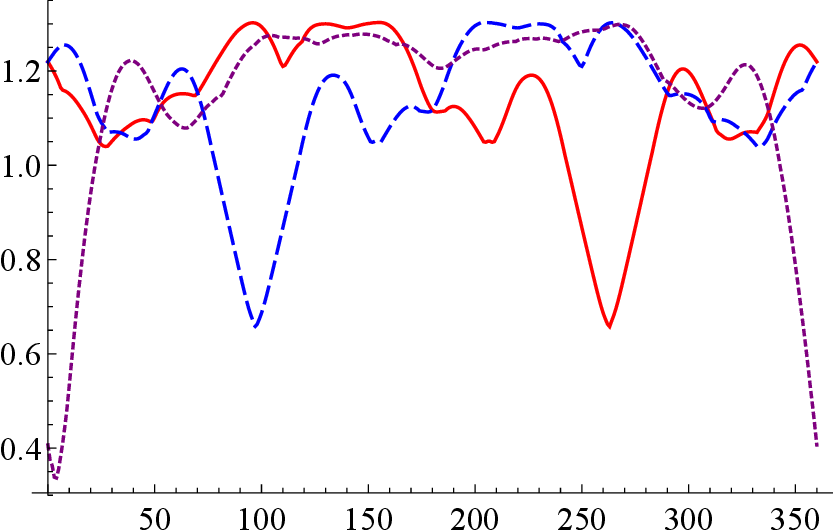}}}		
		\put(5,395){\textit{c}}
		\put(5,172){\textit{d}}
		\put(346,222){$\chi$ [$\null^{\circ}$]}
		\put(20,412){$\Gamma|_{(\mathbf{n}_1,\mathbf{n}_2)}$ [$\mbox{J/m}^2]$}		
		\put(160,286){\textcolor{red}{\large $((1\,1\,1),(2\,1\,1))$}}
		\put(138,380){\textcolor{blue}{\large $((1\,1\,0),(2\,1\,1))$}}	
		\put(157.3,280){\color{red}\line(0,-1){20}}
		\put(150,252){\textcolor{red}{\footnotesize  
				$R([\sqrt{2}\, \overline{1}\, \overline{1}]^T/2,\pi)$}}
		\put(288.5,309){\color{red}\line(0,-1){39}}
		\put(251,261){\textcolor{red}{\footnotesize  
				$R([0\, \overline{1}\, 1]^T/\sqrt{2},\arccos(-\sqrt{8}/3))$}}		
		\put(346,2){$\chi$ [$\null^{\circ}$]}
		\put(20,190){$\Gamma|_{(\mathbf{n}_1,\mathbf{n}_2)}$ [$\mbox{J/m}^2]$}		
		\put(230,60){\large \textcolor{red}{$((1\,2\,3),(3\,5\,7))$}}
		\put(110,60){\textcolor{blue}{\large $((\overline{1}\,\overline{2}\,\overline{3}), (\overline{3}\,\overline{5}\,\overline{7}))$}}
		\put(85,30){\textcolor{purple}{\large $((1\,2\,3), (\overline{3}\,\overline{5}\,\overline{7}))$}}
		\put(263,30){\large\textcolor{purple}{$((1\,2\,3), (\overline{3}\,\overline{5}\,\overline{7}))$}}
	\end{picture}
	\vskip 0.0cm
	\caption{The grain boundary energy of Ni approximated by BRK model
		versus $\chi$
		for selected $\mathbf{n}_1$ and $\mathbf{n}_2$.
		Misorientations for some characteristic points are marked. 
		The plots in  (\textit{a}) have the property
		that the point at $\chi=0$  corresponds to the misorientation $\Sigma1$.
		The inset 
		in (\textit{c}) shows the inverted cusp in magnification.
		Graphs in (\textit{d})
		illustrate the relationship (\ref{eq:equiv_inversion})
		and lack of equivalence between $(\mathbf{n}_1,\mathbf{n}_2)$ and 
		$(\mathbf{n}_1,-\mathbf{n}_2)$. 
	}
	\label{Fig_energy_1D}
\end{figure}

More detailed information is contained in 
plots of $\Gamma|_{(\mathbf{n}_1,\mathbf{n}_2)}$ versus $\chi$.
In the case of twist boundaries, such plots are closely related to frequently drawn 
plots of energy versus misorientation angle.
In contrast, 
considerations regarding
the dependence of the energy on the angle of rotation about axis 
perpendicular to the boundary plane for non-twist boundaries
are difficult to find. 
Example plots of $\Gamma|_{(\mathbf{n}_1,\mathbf{n}_2)}$ versus $\chi$ are shown in 
Fig.~\ref{Fig_energy_1D}. 
To make their symmetries clear, all figures are drawn in the domain from 0 to $2 \pi$.
The functions in Fig.~\ref{Fig_energy_1D}\textit{a},\textit{b},\textit{c}
have 
cusps, i.e., local minima at which 
$\Gamma|_{(\mathbf{n}_1,\mathbf{n}_2)}$ is not differentiable.
In this regard, one needs to note that 
singularities in continuous energy models should arise from physical premises.
Similarly to the case of surface energy and Wulff construction,
there is a physical reason (related to boundary faceting) for
cusps in energy as a function of vectors normal to the boundary plane
\cite{Herring_1951}. 
However, the angle $\chi$ is independent of both $\mathbf{n}_1$ and $\mathbf{n}_2$,  
and if one goes beyond $\Sigma 1$ and the twin boundary, 
there are no simple theoretical arguments for cusps 
in the dependence of $\Gamma|_{(\mathbf{n}_1,\mathbf{n}_2)}$ on $\chi$
and most of these functions could be free of singularities.

\begin{figure}
	\begin{picture}(300,590)(0,0)		
		\put(60,0){\resizebox{10.0 cm}{!}{\includegraphics{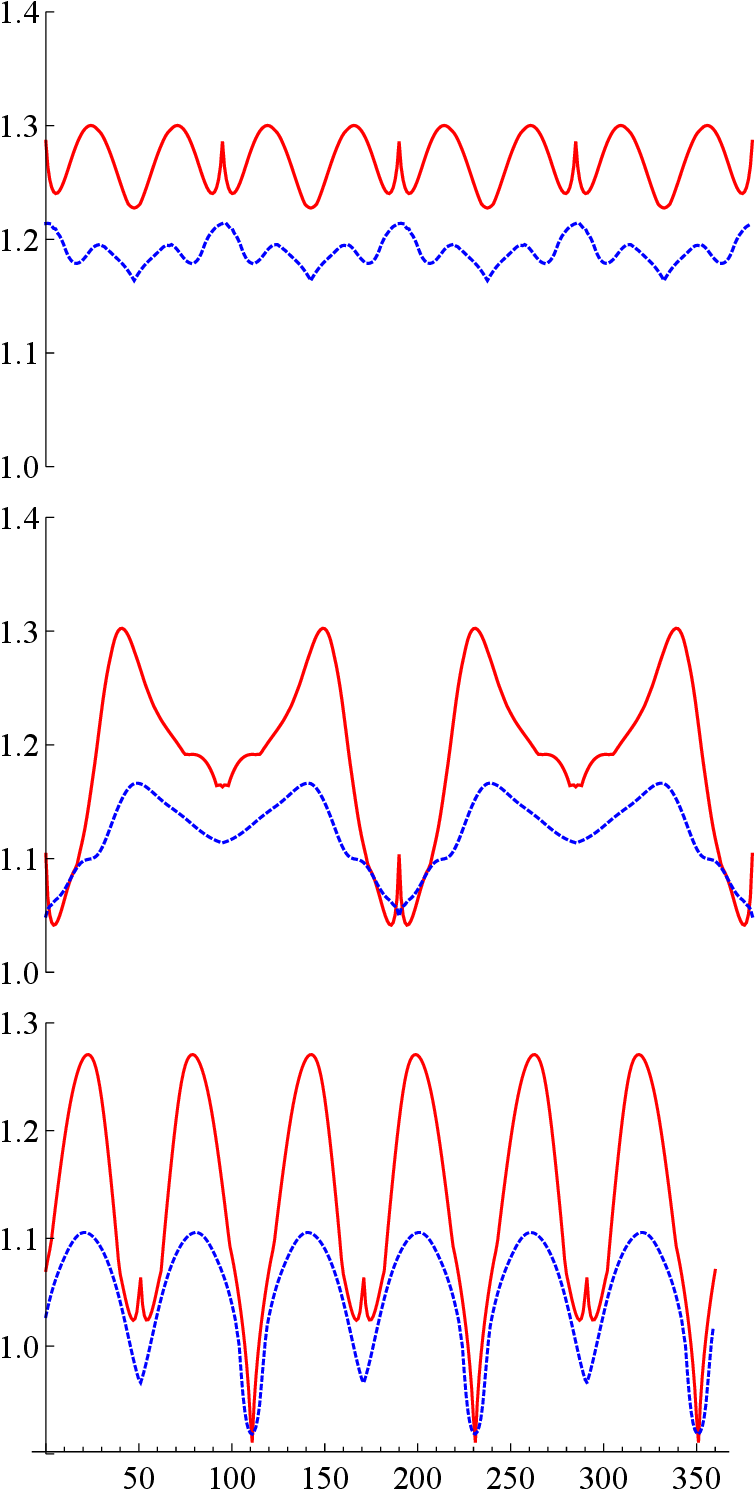}}}
		
		\put(20,571){$\Gamma|_{(\mathbf{n}_1,\mathbf{n}_2)}$ [$\mbox{J/m}^2]$}
		
		\put(167,530){\large $((1\,0\,0),(2\,1\,0))$}
		\put(167,336){\large $((1\,1\,0),(2\,2\,1))$}
		\put(167,174){\large $((1\,1\,1),(2\,1\,1))$}
		\put(346,2){$\chi$ [$\null^{\circ}$]}
		
	\end{picture}
	\vskip 0.0cm
	\caption{Selected BRK-based plots of $\Gamma|_{(\mathbf{n}_1,\mathbf{n}_2)}$ 
		with inverted cusps (red, continuous)
		compared with corresponding data based on the model described in \cite{Morawiec_2025} (blue, dotted).
	}
	\label{Fig_inverted_cusps}
\end{figure}

Interestingly,  the plot for 
$(\mathbf{n}_1,\mathbf{n}_2) \sim ((1\,1\,1),(2\,1\,1))$ in 
Fig.~\ref{Fig_energy_1D}\textit{c} has an inverted cusp,
i.e., a singular point which is a local maximum. 
Of the pairs $(\mathbf{n}_1,\mathbf{n}_2)$ listed in Tab.~\ref*{table:1}, 
inverted cusps are also in the plots for 
$((1\,0\,0),(2\,1\,0))$,
$((1\,1\,0),(2\,1\,0))$,
$((1\,1\,0),(2\,2\,1))$ and
$((1\,1\,1),(2\,2\,1))$.
This raises intriguing questions.
Do the inverted cusps reflect properties of the input data or are due to other causes?
Do other energy models constructed based on the same principles as BRK  
(e.g., \cite{dette2017efficient,sarochawikasit2021grain,chirayutthanasak2022anisotropic,	
chirayutthanasak2024universal}) have inverted cusps?
How would inverted cusps affect physical properties of boundaries? 
What is the impact of inverted cusps on BRK-based grain growth simulations; see, e.g.,
\cite{hallberg2019modeling,nino2023influence,nino2024evolution}?
Resolving these issues is beyond the scope of this study, 
but it is worth noting that the model described in \cite{Morawiec_2025} 
(which, by definition, has no singularities in the dependence on $\chi$)
shows local minima at positions of the  BRK's inverted cusps in 
four out of the five inspected cases,
the only exception being $(\mathbf{n}_1,\mathbf{n}_2) \sim ((1\,1\,0),(2\,1\,0))$;
see, Fig.~\ref{Fig_inverted_cusps}. 
This seems to indicate that the presence 
of inverted cusps is an unintended feature of the BRK model.

\begin{figure}
	\begin{picture}(300,570)(0,0)
		\put(40,400){\resizebox{5.6 cm}{!}{\includegraphics{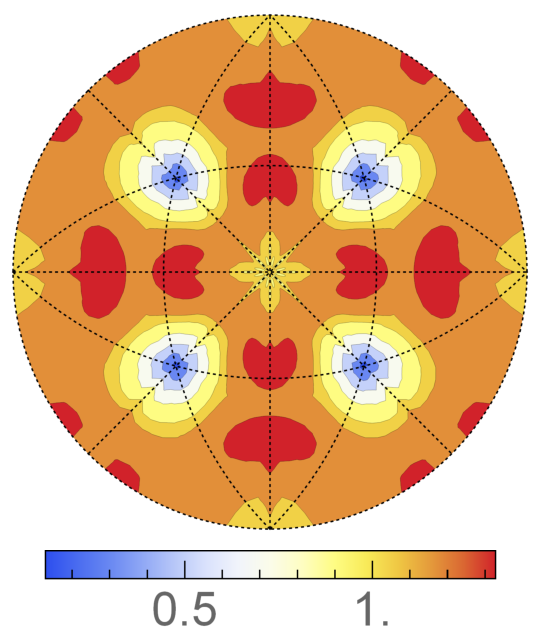}}}
		\put(240,400){\resizebox{5.6 cm}{!}{\includegraphics{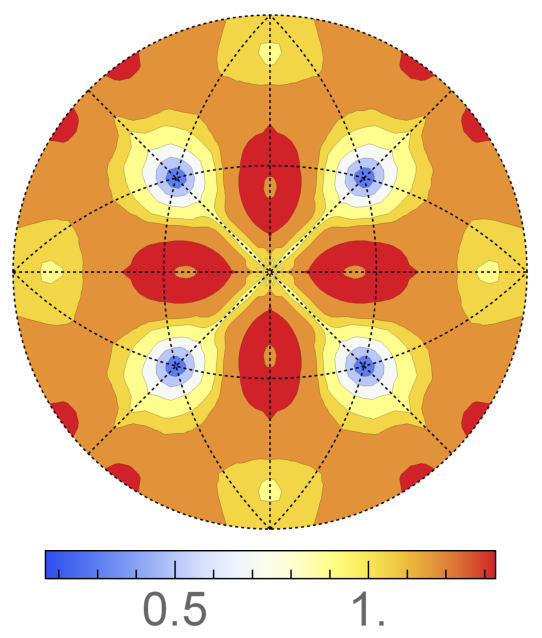}}}
		\put(40,200){\resizebox{5.6 cm}{!}{\includegraphics{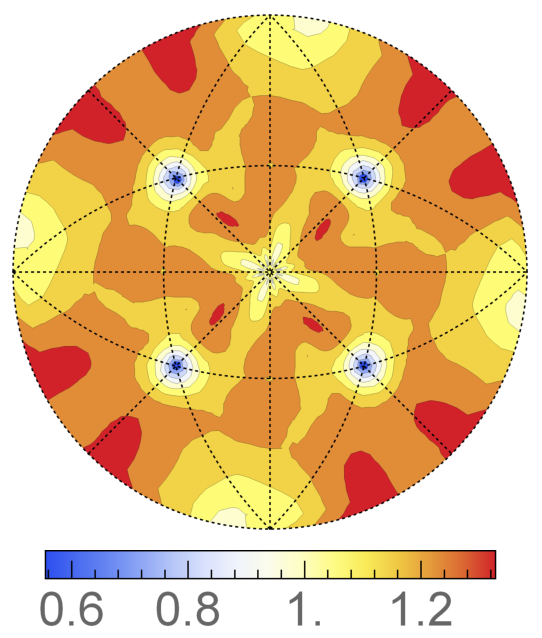}}}
		\put(240,200){\resizebox{5.6 cm}{!}{\includegraphics{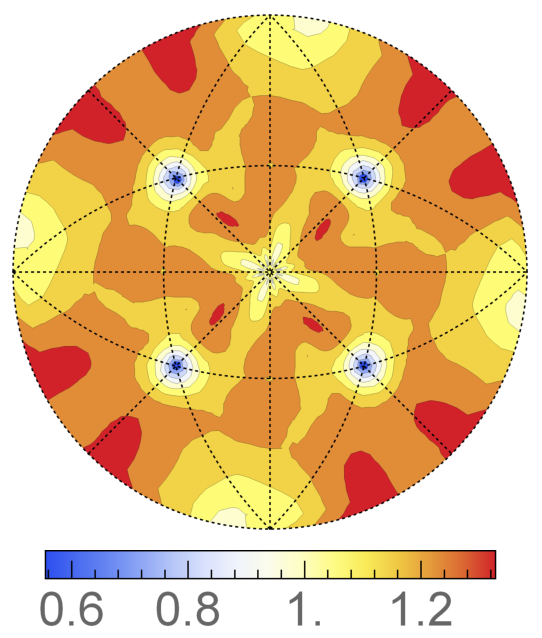}}}
		\put(40,  0){\resizebox{5.6 cm}{!}{\includegraphics{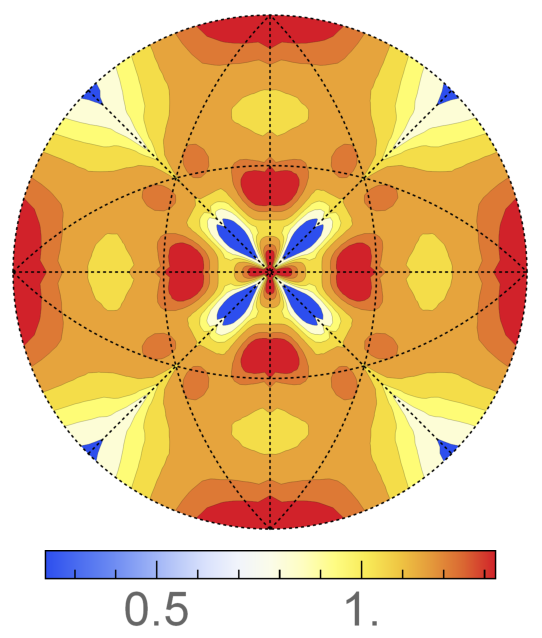}}}		
		\put(240,  0){\resizebox{5.6 cm}{!}{\includegraphics{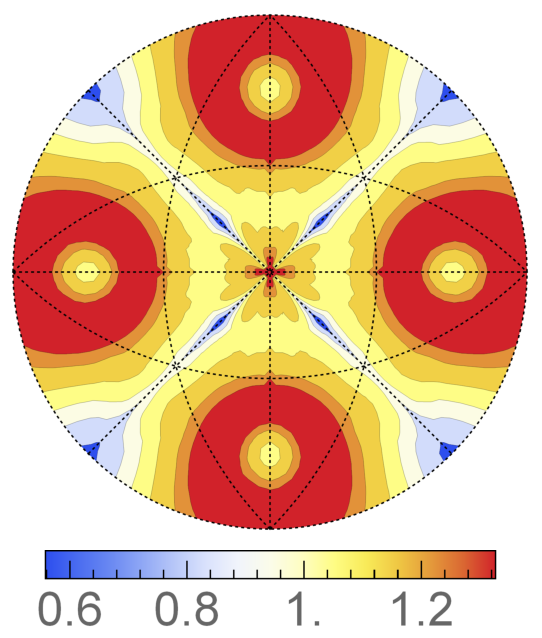}}}		
		\put(20,585){\textit{a}}
		\put(20,385){\textit{b}}
		\put(20,185){\textit{c}}
		\put(170,580){$\mathbf{n}_1 \sim (1\,1\,1)$, $\chi=\pi/3$}
		\put(170,380){$\mathbf{n}_1 \sim (1\,1\,1)$, $\chi=\pi/6$}
		\put(170,180){$\mathbf{n}_1 \sim (1\,1\,4)$, $\chi=\pi$}
		\put(192,550){\vector(-3,-1){45}}
		\put(193,547){\small $\Sigma1$}
		\put(247,550){\vector(3,-1){45}}
		\put(233,547){\small $\Sigma3$}
		\put(190,161){\vector(-3,0){17}}
		\put(190,157){\small $\Sigma3$}
		\put(196,139){\vector(-3,-1){67}}
		\put(195,136){\small $\Sigma1$}
		\put(243,139){\vector(3,-1){67}}
		\put(229,137){\small $\Sigma9$}
		\put(252,149){\vector(2,-1){52}}
		\put(234,149){\small $\Sigma11$}
	\end{picture}
	\vskip 0.0cm
	\caption{
		The grain boundary energy approximation of \cite{bulatov2014grain} for Ni
		versus $\mathbf{n}_2$
		for selected $\mathbf{n}_1$ and $\chi$.
		The plane corresponding to $\mathbf{n}_1$ 
		and the angle $\chi$ are 
		$(1\,1\,1)$ and $\pi/3$ in (\textit{a}),
		$(1\,1\,1)$ and $\pi/6$ in (\textit{b}) and 
		$(1\,1\,4)$ and $\pi$ in (\textit{c}). 
		Left and right columns contain stereographic projections of the upper and
		lower hemishperes from the poles $-\mathbf{e}_z$ and $+\mathbf{e}_z$, respectively.
		Minima of the energy function at some low-$\Sigma$ misorientations are marked. 
		The energy unit is $\mbox{J/m}^2$. The figures were drawn
			using code of G{\l}owi{\'n}ski \cite{Glowinski_2015}. 
	}
	\label{Fig_energy_2D}
\end{figure}

For completeness, it is worth mentioning another 
natural way to visualize $\Gamma$ by showing its dependence 
on one of the vectors $\mathbf{n}_i$ when the other vector 
and $\chi$ are fixed.
Such sections through $\Gamma$ for selected 
$\mathbf{n}_1$ and $\chi$ and variable $\mathbf{n}_2$ are shown in Fig.~\ref{Fig_energy_2D}.
They resemble graphs plotted for a constant misorientation and variable 
$\mathbf{n}_1$
(see, e.g., \cite{saylor2003relative}), 
but their interpretation is more complicated. 
The four-fold symmetry about $\mathbf{e}_z$ visible in Fig.~\ref{Fig_energy_2D} 
is the direct result of (\ref{eq:equiv_CBC}).
The function 
shown in 
Fig.~\ref{Fig_energy_2D}\textit{b}  
is also symmetric with respect the change of sign of $\mathbf{n}_2$.

It has been suggested that properties of boundaries in cubic metals 
are controlled more by the boundary planes rather than misorientation;
see, e.g., \cite{Randle_2006,Rohrer_2011,Ratanaphan_2015}.
This claim 
is partly obscured by the fact that a change 
of one crystal plane with the other fixed 
is possible only by a change of misorientation.
Since the angle $\chi$ is independent of both $\mathbf{n}_1$ and $\mathbf{n}_2$, 
the interface-plane parameterization provides an opportunity to overcome this obscurity.
The dimension of the subspace determined by $\mathbf{n}_1$ and $\mathbf{n}_2$, which is four, 
is large compared to the dimension of the subspace defined by $\chi$, which is one.
This observation alone may lead to the conclusion 
that the influence of boundary planes is greater 
than the influence of misorientation.
On the other hand, 
the level of 
control of a property by misorientation can be compared to that by boundary planes 
by looking at two distinct cases:
one, with fixed $\mathbf{n}_i$ $(i=1,2)$ and variable $\chi$, and second, 
with fixed misorientation $M$ and variable angle, say $\zeta$, 
representing a change of $\mathbf{n}_1$. 
One may assume that 
the impact of a given parameter on a function
is reflected in amplitude and frequency of changes in the function:
the higher the impact, the larger and more frequent the changes.
If the influence of the boundary plane on the boundary energy 
were larger than the influence of misorientation, 
the variation of the energy function over $\zeta$ 
would be larger than the variation of the function over $\chi$.
A comparison of standard deviations from the means for twenty random energy 
plots with respect to each of $\chi$ and $\zeta$ shows the opposite.
The standard deviations
for the energy functions over $\chi$
were found to be on average more than twice as large as 
the deviations for the distributions over  $\zeta$.
In other words, the comparison of variations of the energy over 
$\chi$ and $\zeta$ does not confirm the suggestion 
mentioned at the beginning of this paragraph.

\section{Concluding remarks}

Macroscopic parameters are an important tool in research on grain boundaries
and intercrystalline interfaces in general. 
Such parameters are not universal. 
Parameters useful in some circumstances may be inconvenient in other cases.
Those facilitating interpersonal communication 
need to be easy to interpret and appeal to physical intuition.
Using several different boundary parameterizations 
allows for looking at boundaries and functions defined on the boundary space 
from multiple perspectives.

The `interface-plane parameterization' introduced in this paper is based on  
two vectors $\mathbf{n}_i$ ($i=1,2$) identifying the orientation 
of the boundary plane in crystals on both sides of the boundary
and the angle $\chi$ of rotation about the axis perpendicular to the plane, 
i.e., a boundary is represented by the triplet $(\mathbf{n}_1,\chi,\mathbf{n}_2)$. 
The triplet is directly linked to Euler angles representing 
crystal orientations. 
In many practical cases, the vectors $\mathbf{n}_i$ ($i=1,2$) can be replaced by 
Miller indices.

The characterization of a boundary by the crystallographic planes
is natural and intuitive.  
Also the rotation by the angle $\chi$ can be easily visualized, 
but capturing the configuration with $\chi=0$ is generally complicated.
Apart from simple special cases, it requires numerical calculations.

The interface-plane parameters of some characteristic boundary types are easy to interpret.
This concerns symmetric boundaries represented by $(\mathbf{n},\pi,-\mathbf{n})$,
twist boundaries represented by $(\mathbf{n},\chi,-\mathbf{n})$,
and  $180^{\circ}$-tilt boundaries represented by $(\mathbf{n},\chi,\mathbf{n})$.
However, the representation of general tilt boundaries is more involved.

An examination of the prominent Bulatov-Reed-Kumar \cite{bulatov2014grain}
grain boundary energy model as a function of 
the interface-plane parameters revealed its previously unknown features. 
The energy dependence on $\chi$ 
shows inverted cusps for some pairs $(\mathbf{n}_1,\mathbf{n}_2)$. 
Standard deviations from the means for energy versus $\chi$ 
were found to be larger than for energy versus 
boundary plane changes, which is contrary to the 
assertion that boundary properties are controlled more by 
boundary planes than by misorientations.

For simplicity, 
the paper concerns homophase boundaries between centrosymmetric crystals.
However, from the viewpoint of macroscopic description, 
the cases of homo- and hetero-phase interfaces 
between centrosymmetric crystals
differ in non-essential aspects. 
(There is no heterophase analogue to the `no-boundary' case, 
and the exchange symmetry is forbidden.)
Therefore, the scope of the described parameterization 
can be easily extended to interfaces 
between centrosymmetric crystals of different types.

There are no fundamental obstacles to abandoning 
the assumption of centrosymmetry and generalizing 
the interface-plane parameterization to interfaces 
between crystals with arbitrary point groups.
With such a general approach, 
it would be necessary to separately consider the cases of centrosymmetric crystals, 
acentric but non-enantiomorphic crystals, and enantiomorphic crystals, 
as well as their combinations if inter-phase interfaces are taken into account.

\vskip 1.0cm

\subsection*{Appendix A: Interface-plane parameters from interface matrix}

The triplet $(\mathbf{n}_1, \chi ,\mathbf{n}_2)$ or the parameters
$\theta_1,\psi_1, \,  \chi , \,  \theta_2,\psi_2$ 
can be calculated from the interface matrix $\widetilde{\mathbf{B}}=\mathbf{B}(M,\mathbf{n}_1)$ 
using elementary steps.
The vectors $\mathbf{n}_1$ and $\mathbf{n}_2$ 
are directly accessible as parts of this matrix.
The angles $\theta_i$ and $\psi_i$
are obtained from $\mathbf{n}_i$ ($i=1,2$) using 
\bEq
\begin{array}{l}
	(\theta_1,\pi/2-\psi_1)=\Theta^{-1}(+\mathbf{n}_1) \ \mbox{and} \ \ 
	(\theta_2,\pi/2-\psi_2)=\Theta^{-1}(-\mathbf{n}_2) \ ,
\end{array} 
\label{eq:inv_Theta}
\eEq
where 
$\Theta^{-1}$ denotes the function 
inverse to $\Theta$ given in (\ref{eq:vfromspherical}). Explicit expressions for  
spherical coordinates $(\theta,\varphi)=\Theta^{-1}(\mathbf{n})$ of the 
unit vector $\mathbf{n}=[n^1 \, n^2 \, n^3  ]^T$ are 
$\theta = \arccos (n^3)$ and 
$\varphi = \arctan(n^1,  n^2)$ if $|n^3| \neq 1$ and $\varphi = 0$ 
otherwise.\footnote{Two-argument 
$\arctan$ used here follows the convention that for $c > 0$ 
one has $\arctan(c \cos \alpha, c \sin \alpha) = \alpha$.}

As for the determination of the angle $\chi$, based on eq.(\ref{eq:basic_relationship}), 
one has the relationship
$\mathbf{Q}(\theta_1,\psi_1)^T  \widetilde{\mathbf{B}}  \mathbf{Q}(\theta_2,\psi_2)
=\mathbf{B}( R(\mathbf{e}_z, \chi),\mathbf{e}_z)$
or 
$Q(\theta_1,\psi_1)^T M Q(\theta_2,\psi_2) = 
R(\mathbf{e}_z  ,\chi) = P$, and hence, 
knowing $\theta_i$, $\psi_i$ and $M$,
the angle is obtained via
$\chi  = \arctan(P_{11},  P_{12})$. 
Alternatively, the angle $ \chi $ can be computed directly from the 
entries of $\widetilde{\mathbf{B}} = \mathbf{B}(M,\mathbf{n}_1)$ 
without referring to $\theta_i$ and $\psi_i$ by using
$$
\begin{array}{rcl}
	\chi   & = & 
	\arctan(n_1^3 n_2^3 + M_{33},  \ M_{23}  n_1^1 - M_{13}  n_1^2)  \ \mbox{or} \\
	\chi  & = & 
	\arctan(n_1^3 n_2^3 + M_{33},  \ M_{32} n_2^1 - M_{31} n_2^2)  \ , \\
\end{array}
$$
where $n_i^k$ is the $k$-th component of $\mathbf{n}_i$.
The above relationships are applicable if $| n_i^3 | \neq 1$. 
In the remaining cases, one has 
$$
\chi  =\left\{
\begin{array}{ll}
	\arctan(\mp M_{13} , \ +M_{23})	
	& \mbox{if}  \ 
	n_1^3=\pm 1 \ \mbox{and}  \  | n_2^3 | \neq  1 \ , 
	\\
	\arctan(\pm M_{31} , \ -M_{32} )
	& \mbox{if}  \ 
	n_2^3=\pm 1 \ \mbox{and}  \ | n_1^3 | \neq  1 \ , 
	\\
	\arctan(\pm M_{11} , \ -M_{12} )
	& \mbox{if}  \ 
	n_1^3= - 1 \ \mbox{and}  \  n_2^3= \pm 1 \ , 
	\\
	\arctan(\mp M_{11} , \ +M_{12} )
	& \mbox{if}  \ 
	n_1^3= + 1 \ \mbox{and}  \  n_2^3= \pm 1 \ , 
\end{array}
\right.
$$
where either the lower or the upper signs are used concurrently.

\subsection*{Appendix B: Condition for tilt boundaries}

In the case of a tilt boundary, 
the misorientation axis $\mathbf{k}$ is perpendicular to $\mathbf{n}_1$, i.e., 
$\mathbf{k} \cdot \mathbf{n}_1=0$. 
(This implies that also $\mathbf{k} \cdot \mathbf{n}_2=0$.)
In terms of the parameters 
$\left(\theta_1, \psi_1, \chi , \theta_2, \psi_2\right)$, this condition
has the form
\bEq
q_c \cos \chi + q_s \sin \chi = 0 \ ,
\label{eq:general_tilt}
\eEq
where
$$
\begin{array}{l}
	q_c = \sin \left(\psi_1-\psi_2\right) \left(\cos \theta_1 + \cos \theta_2\right) \ ,
	\\
	q_s = \cos \left(\psi_1-\psi_2\right) \left(\cos \theta_1 \cos \theta_2+1\right)
	+\sin \theta_1 \sin \theta_2  \ . 
\end{array}
$$
Using the relationships (\ref{eq:inv_Theta})
between $(\theta_i, \psi_i)$ and $\mathbf{n}_i$, 
one can express $q_c$ and $q_s$ through components of $\mathbf{n}_i$.
If eq.(\ref{eq:general_tilt}) is satisfied, the misorientation axis is in the boundary plane
or the misorientation angle is zero.

\clearpage

\bibliographystyle{unsrt}
\bibliography{GB_Matrix.bib} 

\end{document}